\documentclass{article}

\usepackage{arxiv}

\usepackage[utf8]{inputenc} 
\usepackage{graphicx}
\usepackage{cite}
\usepackage[T1]{fontenc}    
\usepackage{hyperref}       
\usepackage{url}            
\usepackage{booktabs}       
\usepackage{amsfonts}       
\usepackage{nicefrac}       
\usepackage{microtype}      
\usepackage{lipsum}		
\usepackage{graphicx}
\usepackage{doi}

\makeatletter
\def\@cite#1{\textsuperscript{[#1]}}
\makeatother

\title{SciQu: Accelerating Materials Properties Prediction with Automated Literature Mining for Self-Driving Laboratories}

\author{ \href{https://orcid.org/my-orcid?orcid=0000-0003-4746-6011}{\includegraphics[scale=0.06]{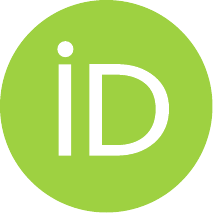}\hspace{1mm}Anand Babu} \\
	Quantum Materials and Devices Unit\\
	Institute of Nano Science and Technology\\
	Knowledge City, Sector 81, Mohali 140306, India \\
	\texttt{ababu.nano@gmail.com} \\
}

\date{}

\renewcommand{\headeright}{}
\renewcommand{\undertitle}{}
\renewcommand{\shorttitle}{}

\hypersetup{
pdftitle={SciQu: Accelerating Materials Properties Prediction with Automated Literature Mining for Self-Driving Laboratories},
pdfsubject={q-bio.NC, q-bio.QM},
pdfauthor={Anand Babu},
pdfkeywords={Materials properties, Machine learning, Literature mining, Self-driving laboratories},
}
\begin{document}
\maketitle

\begin{abstract}
	Assessing different material properties to predict specific attributes, such as band gap, resistivity, young modulus, work function, and refractive index, is a fundamental requirement for materials science-based applications. However, the process is time-consuming and often requires extensive literature reviews and numerous experiments. Our study addresses these challenges by leveraging machine learning to analyze material properties with greater precision and efficiency. By automating the data extraction process and using the extracted information to train machine learning models, our developed model, SciQu, optimizes material properties. As a proof of concept, we predicted the refractive index of materials using data extracted from numerous research articles with SciQu, considering input descriptors such as space group, volume, and bandgap with Root Mean Square Error (RMSE) $\sim$ 0.068 and $R^{2}$ $\sim$ 0.94. Thus, SciQu not only predicts the material's properties but also plays a key role in self-driving laboratories by optimizing the synthesis parameters to achieve precise shape, size, and phase of the materials subjected to the input parameters.
\end{abstract}


\section{Introduction}
The rapid expansion of materials science has led to an overwhelming amount of scientific research and publications. However, the sheer volume of literature presents significant challenges for researchers needing to understand the properties of the materials in the rapidly advancing field of materials science. \cite{ref1, ref2, ref3} Traditional manual literature reviews are increasingly impractical, particularly in predicting specific attributes of materials, such as band gap, resistivity, Young's modulus, work function, and refractive index, and optimizing parameters, including precursor materials, solvents, reaction conditions, and post-synthesis treatments in the synthesis of materials. \cite{ref3, ref4} 

Researchers typically rely on exhaustive literature reviews to identify optimal properties of the materials and their synthesis parameters, a process that is not only time-consuming but also susceptible to human error and bias. There is a pressing need for automated tools that streamline the literature review process and provide reliable, data-driven recommendations. \cite{ref5,ref6} Artificial intelligence (AI) and natural language processing (NLP) technologies offer promising solutions in this context. Large Language Models (LLMs) such as Bidirectional Encoder Representations from Transformers (BERT) and different versions of Generative Pre-trained Transformer (GPT) have shown remarkable capabilities in text summarization, question answering, and information retrieval. These models can automate the reading and analysis of scientific literature, improving research efficiency and accuracy .\cite{ref7, ref8, ref9}

Despite the significant advancements in LLMs, existing tools often struggle with the complexity and specificity of scientific language. Scientific articles typically contain dense, technical jargon and intricate data representations, making it challenging for generic NLP models to accurately extract relevant information. \cite{ref10, ref11, ref12, ref13, ref14, ref15} Moreover, traditional LLMs are not inherently designed to handle the unique requirements of scientific research, such as the need to cross-reference multiple sources and maintain context over long documents. Addressing these limitations requires the development of specialized tools that integrate advanced retrieval techniques and custom prompt systems tailored to scientific content. \cite{ref16, ref17, ref18, ref19, ref20} 
\begin{figure}[htbp]%
\centering
\includegraphics[width=0.8\textwidth]{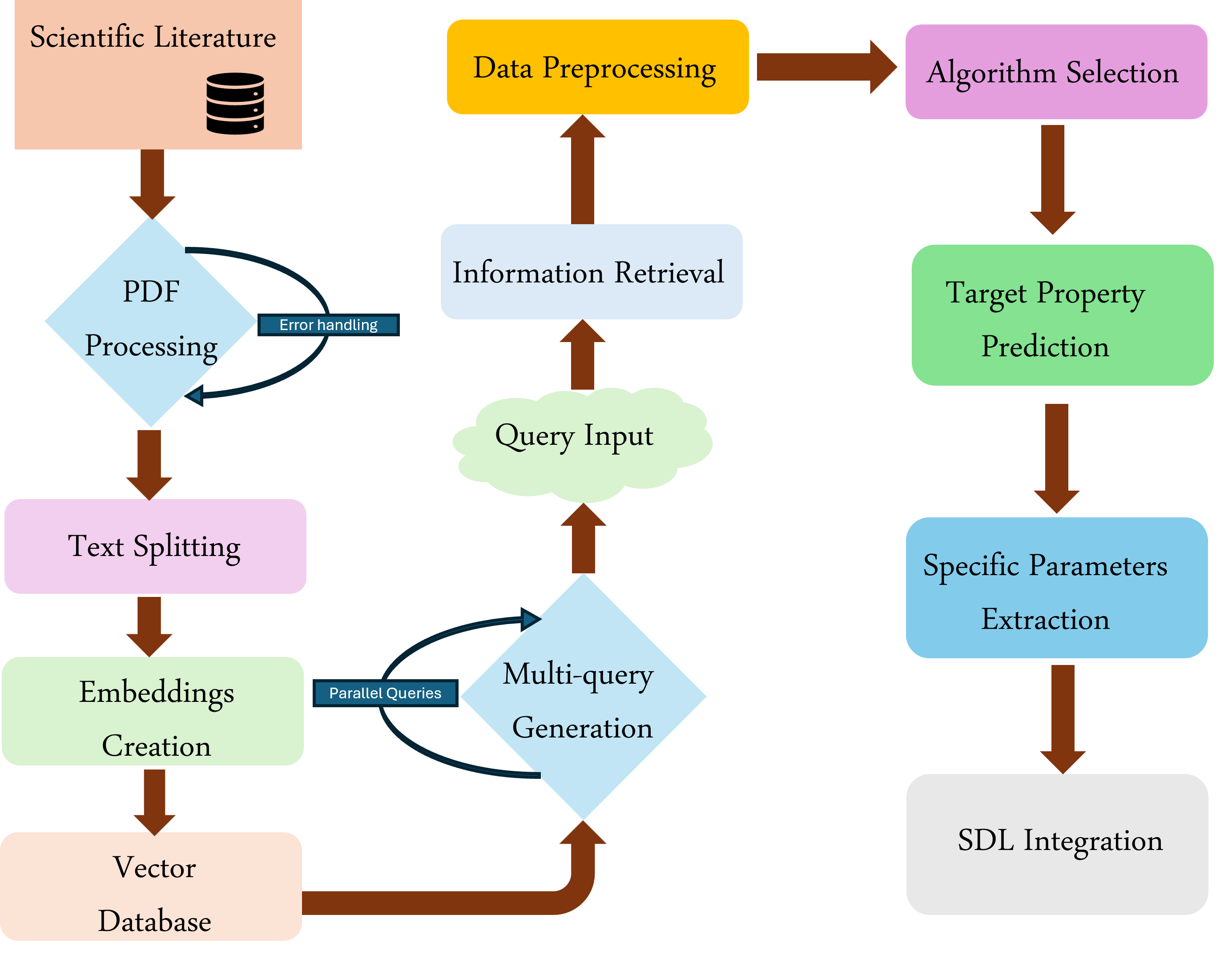}
\caption{Workflow of the SciQu for various materials property prediction autonomously. The process includes data extraction from a scientific database, text processing and embedding creation, efficient information retrieval, and algorithmic prediction of target properties. The optimized parameters are then integrated into self-driving laboratories (SDLs).}
\label{fig:Fig1}
\end{figure}
Several specialized tools have been developed to address the unique challenges of scientific literature review. For instance, the Semantic Scholar Open Research Corpus (S2ORC) dataset offers a vast collection of scientific articles with structured metadata, enhancing information retrieval and analysis. \cite{ref21} Similarly, the SciBERT model, a variant of BERT trained specifically on scientific texts, has shown improved performance in tasks like entity recognition and relation extraction in scientific documents. \cite{ref22} AlloyBERT has been employed for predicting alloy material properties \cite{ref23}, and LittLLM has been used for comprehensive scientific literature reviews. \cite{ref24, ref25} Despite these advancements, a significant gap remains in developing tools that can seamlessly integrate literature review with experimental optimization and automation, often due to challenges such as introducing technical jargon, hallucinations, and misunderstandings of scientific terminology.

In this context, we introduce SciQu, tool that addresses these challenges by extracting key scientific information from literature articles and leveraging machine learning to predict the new properties of the materials. It enhances the efficiency and accuracy of the synthesis process by combining literature review with automated experimental optimization. SciQu automates data extraction and uses the extracted information to train machine learning models, providing the complex relationship among the different properties of the materials. As a pilot demonstration, we have predicted the refractive index of the materials, a crucial parameter for optoelectronic device application, extracting from the various articles where bandgap, space group, volume, no. of sites, and other properties have been utilized as input. Thus, our proposed approach not only accelerates the research process but also significantly enhances the accuracy and reproducibility of experimental outcomes. The development of SciQu represents a significant advancement in the application of AI and NLP technologies to scientific research.

\section{Methodology}
\label{sec:headings}
The flowchart illustrates the methodology of SciQu system designed to predict the properties of materials (Figure \ref{fig:Fig1}). This comprehensive approach integrates several key steps, from data preprocessing to algorithmic prediction.The methodology begins with accessing scientific literature relevant to various material properties such as band gap, refractive index, space group, work function, and others. PDF documents from the database undergo processing to extract textual information while incorporating robust error-handling mechanisms to ensure data integrity.
\begin{figure}[htbp]%
\centering
\includegraphics[width=1\textwidth]{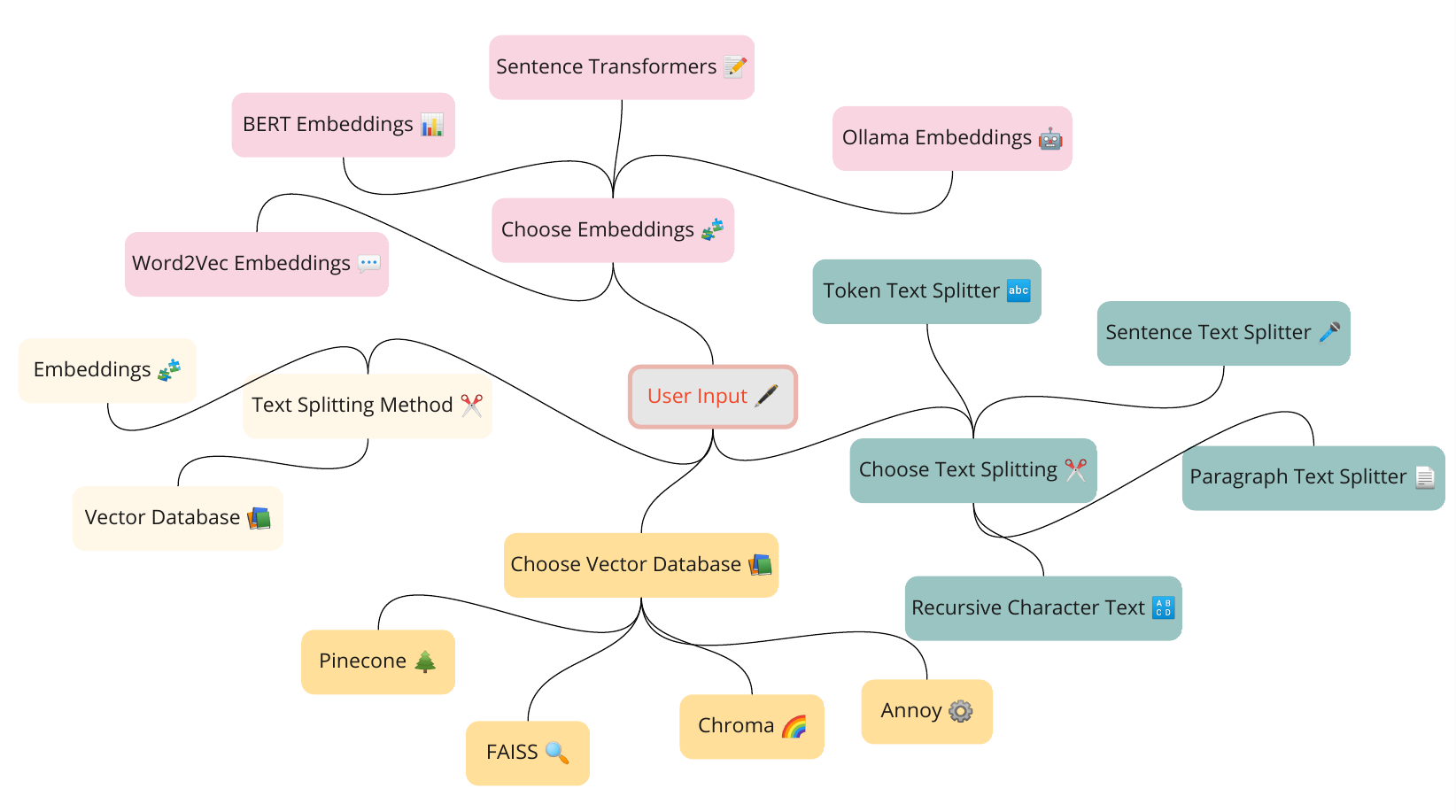}
\caption{Different modules and libraries for the customized SciQu model as per the requirement.}
\label{fig:Fig2}
\end{figure}
 Text splitting techniques are then employed to divide the extracted text into manageable segments, facilitating detailed analysis. Once the text is split, it is converted into vector embeddings, which capture the semantic meaning of the text and are stored in a vector database. The vector database enables efficient retrieval of relevant information, a crucial component for subsequent data analysis and query handling. User queries initiate the multi-query generation process, that allows for the parallel execution of multiple queries to the vector database. This parallel processing enhances the speed and efficiency of information retrieval.

 The retrieved information is then used to address the specific needs of the query, ensuring that relevant data is promptly available for analysis. The retrieved data undergoes further preprocessing to prepare it for algorithmic analysis. This step is critical as it ensures that the data is clean, well-structured, and suitable for machine learning models. The choice of algorithm is based on the specific characteristics of the data and the desired outcomes.  Using the selected algorithm, the system predicts the target properties of the materials. Besides the materials property prediction, SciQu has the potential to optimize several key parameters for the synthesis of specific materials with the desired properties.
To ensure SciQu delivers robust and reproducible results, various functionalities have been employed, including diverse text-splitting methods and multiple data embeddings. Figure \ref{fig:Fig2} illustrates the core functionalities and workflow of SciQu. Predicting the various properties of materials involves optimizing a multitude of parameters, a traditionally labor-intensive process requiring comprehensive literature reviews and numerous experimental iterations. SciQu addresses these challenges by automating data extraction and parameter optimization, thereby accelerating and refining the synthesis workflow. The following sections detail the alternative tools and functionalities used to build an efficient and accurate model.

\subsection{Automated Data Extraction}
SciQu leverages advanced text processing techniques to extract relevant scientific information from literature articles. The diagram outlines the critical steps involved in this process, beginning with user input and subsequent text-splitting methods. The use of various text splitters—Paragraph, Token, and Sentence Text Splitters—ensures that the text data is divided into appropriately sized chunks for efficient processing. This segmentation is essential for the accurate extraction of contextually rich information, which forms the basis for training machine learning models. \cite{ref26, ref27, ref28}

\subsection{Embedding Techniques}
The extracted text chunks are transformed into vector representations using multiple embedding techniques such as Word2Vec, BERT, Sentence Transformers, and Ollama Embeddings. These embeddings capture the semantic relationships and contextual nuances of the text, making them highly valuable for downstream machine-learning tasks. \cite{ref29, ref30} In SciQu, these embeddings are crucial for training models that predict the various materials properties and optimized synthesis parameters. The choice of embedding technique can significantly influence the model's performance, and SciQu’s flexibility in using different embeddings ensures robust and accurate predictions.

\subsection{Vector Databases}
The vectorized data is stored in vector databases like Facebook AI Similarity Search (FAISS), Chroma, Pinecone, and Annoy. \cite{ref31, ref32, ref33} These databases are optimized for high-dimensional similarity search, enabling efficient retrieval of relevant information. In the context of SciQu, these databases facilitate quick access to scientific data and support the iterative process of model training and descriptors optimization. 

\section{Results and Discussion}
\begin{figure}[htbp]%
\centering
\includegraphics[width=0.8\textwidth]{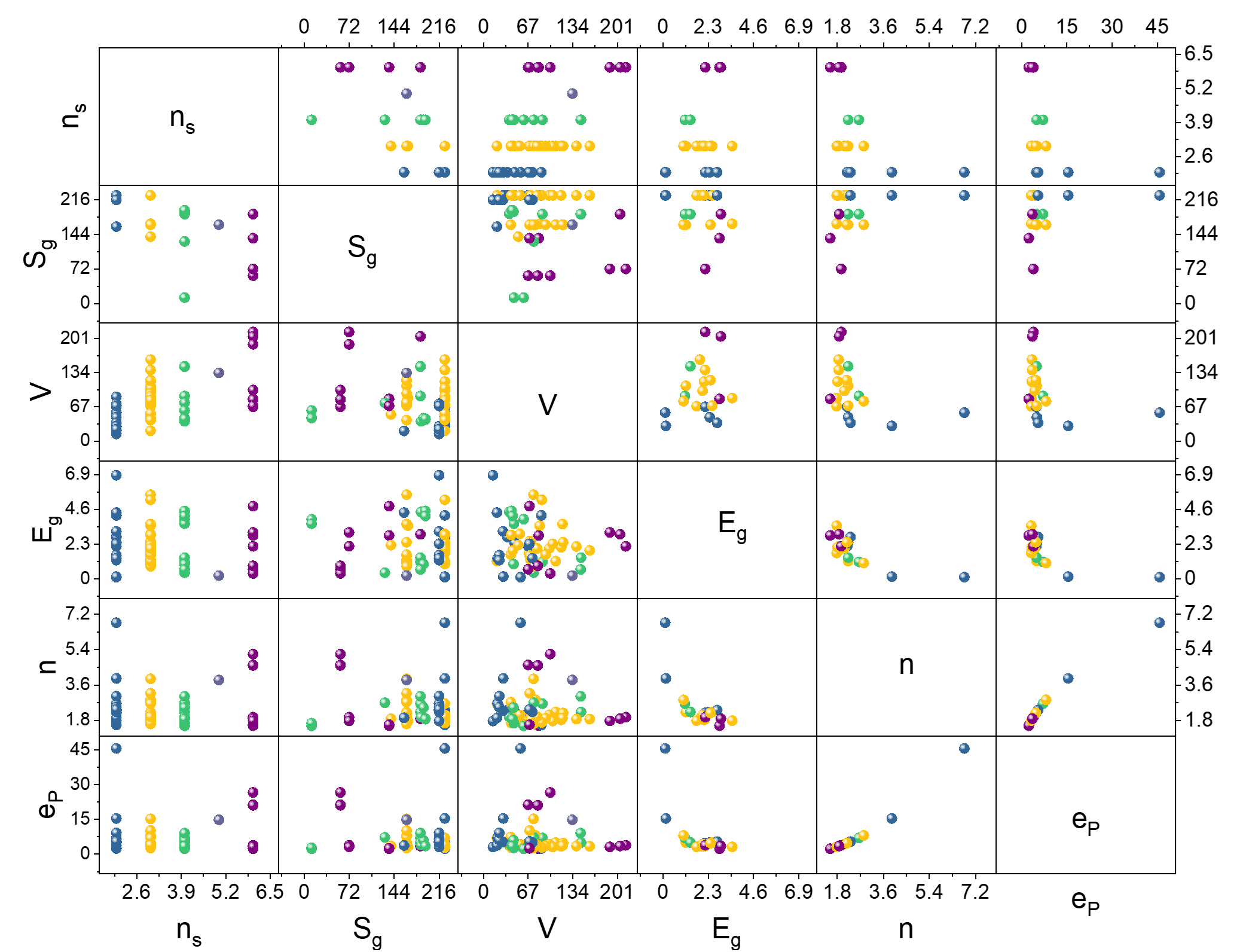}
\caption{Scatterplot matrix visualizing the relationships among various material properties.}
\label{fig:Fig3}
\end{figure}

Various properties of the materials, such as volume (V), band gap ($E_{g}$), number of sites ($n_{s}$), space group ($S_{g}$), refractive index (n), and polyelectronic property ($e_{P}$) have been extracted from various research articles (Table \ref{table:Table1}). These properties were meticulously analyzed to understand their relationships and impacts on each other. The scatterplot matrix (Figure \ref{fig:Fig3}) provides an overview of these relationships, highlighting potential correlations and anomalies within the dataset. A few outliers, particularly at higher values, indicate potential areas for model improvement and suggest the need for further investigation into the underlying causes of these anomalies. These outliers may be due to specific structural or compositional characteristics of the materials that are not fully captured by the current model.
\begin{table}[htbp]
    \centering
    \vspace{-0.2cm} 
    \caption{Extracted data from the scientific literature.}
    \resizebox{0.65\textwidth}{!}{
    \scriptsize
    \begin{tabular}{llllllllll}
    \hline
        S. No. & $n_{s}$ & $S_{g}$ & V & $E_{g}$ & n & $e_{P}$ & Ref.  \\ \hline
        1 & 3 & 225 & 139.4678 & 2.14 & 1.86 & 3.47 & \cite{ref33}  \\ 
        2 & 3 & 225 & 68.2689 & 1.71 & 1.78 & 3.18 & \cite{ref34}  \\ 
        3 & 2 & 225 & 67.30959 & 2.15 & 2.19 & 4.81 & \cite{ref35}  \\
        4 & 3 & 225 & 98.59723 & 2.02 & 2.08 & 4.32 & \cite{ref36}  \\ 
        5 & 2 & 225 & 55.78457 & 0.12 & 6.76 & 45.7 & \cite{ref37}  \\ 
        6 & 2 & 225 & 46.69512 & 2.38 & 2.23 & 4.99 & \cite{ref38}  \\ 
        7 & 2 & 225 & 35.74349 & 2.76 & 2.32 & 5.4 & \cite{ref39}  \\ 
        8 & 3 & 164 & 119.4609 & 2.42 & 2.18 & 4.73 & \cite{ref40}  \\ 
        9 & 6 & 186 & 204.8833 & 2.94 & 1.88 & 3.54 & \cite{ref41}  \\ 
        10 & 2 & 225 & 29.75755 & 0.15 & 3.93 & 15.43 & \cite{ref42}  \\ 
        11 & 3 & 225 & 45.15779 & 0.97 & 1.96 & 3.84 & \cite{ref43}  \\ 
        12 & 4 & 129 & 75.28545 & 0.41 & 2.7 & 7.29 & \cite{ref44}  \\ 
        13 & 4 & 187 & 38.72896 & 4.42 & 1.96 & 3.84 & \cite{ref45}  \\ 
        14 & 3 & 164 & 40.99221 & 1.6 & 2.73 & 7.44 & \cite{ref46}  \\ 
        15 & 3 & 225 & 103.982 & 2.32 & 1.76 & 3.1 & \cite{ref47}  \\ 
        16 & 2 & 216 & 28.95581 & 3.14 & 2.29 & 5.26 & \cite{ref48}  \\ 
        17 & 3 & 164 & 118.6431 & 3.62 & 1.92 & 3.67 & \cite{ref49}  \\ 
        18 & 2 & 225 & 86.78051 & 4.19 & 1.57 & 2.45 & \cite{ref50}  \\ 
        19 & 3 & 164 & 70.42276 & 1.25 & 1.98 & 3.92 & \cite{ref51}  \\ 
        20 & 3 & 225 & 70.93384 & 2.44 & 1.87 & 3.51 & \cite{ref52}  \\ \hline
    \end{tabular}
    }
    \label{table:Table1}
\end{table}

The scatterplot matrix reveals a clear correlation among different properties. Volume shows weak correlations with other variables like refractive index (n), and polyelectronic property ($e_{P}$), suggesting complex underlying factors that influence these properties independently. The scatterplot of $E_{g}$ versus $n_{s}$ suggests some structural constraints, while $E_{g}$ and $e_{P}$ display dispersed patterns, indicating their dependency on multifaceted intrinsic properties. This dispersion may be attributed to the intricate electronic and structural characteristics inherent to different material systems. \cite{ref53, ref54} The data was further pre-processed to ensure it was suitable for the prediction of the material’s properties, particularly the refractive index (n). 

\begin{figure}[htbp]%
\centering
\includegraphics[width=1\textwidth]{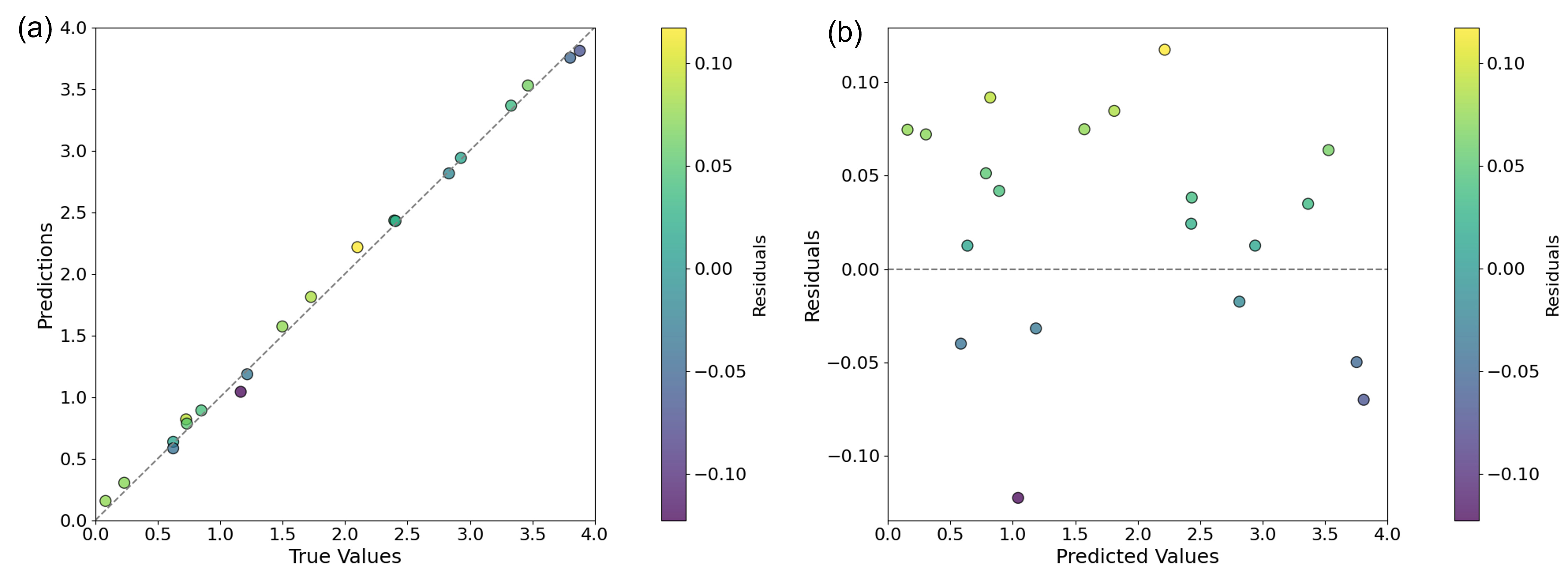}
\caption{Prediction of the refractive index from the extracted data.}
\label{fig:Fig4}
\end{figure}

Random Forest has been utilized for the prediction of the refractive index, with Figure \ref{fig:Fig4}a indicating the prediction line compared with the true values provided in the dataset. The close alignment between predicted and actual values demonstrates the model's accuracy. Figure \ref{fig:Fig4}b displays residuals against predicted values, with residuals centered around zero, suggesting no significant bias and consistent variance in prediction errors. This indicates the model's reliability and robustness in predicting the refractive index across different material systems. The resultant $R^{2}$ value has been found to be approximately 0.94, with the RMSE around 0.068. These metrics indicate a high level of accuracy and precision in the model's predictions.  By leveraging machine learning, automated experimentation, and continuous feedback, SciQu not only accelerates the research workflow but also ensures that the synthesized materials meet precise specifications. This capability demonstrates SciQu's potential to revolutionize materials research and self-driving laboratories by providing rapid, accurate predictions that guide experimental efforts efficiently. The use of SciQu in this study highlights its effectiveness in mining scientific literature to extract relevant data, train predictive models, and apply these models to real-world materials synthesis problems. The integration of machine learning with automated experimentation creates a feedback loop that continuously improves the accuracy of predictions and the quality of synthesized materials.
This approach represents a significant advancement in the field of materials science, enabling researchers to achieve their objectives more quickly and with greater precision than traditional methods. Furthermore, the correlations and patterns identified in this study provide valuable insights into the fundamental relationships between material properties and their structural characteristics. \cite{ref55, ref56} These insights can guide future research efforts, helping to identify new materials with desirable properties and optimize synthesis processes for existing materials.

The ability to predict material properties accurately also has important implications for the development of new technologies and applications, as it allows researchers to tailor materials to specific needs and performance criteria. Besides the new property protection, SciQu has the potential to integrate seamlessly with the SDLs. The process begins with predicted parameters derived from extensive data analysis and machine learning models. These parameters are then interfaced with microcontrollers to direct the experimental setup. Automated experimentation follows, where the microcontroller executes the experimental protocol based on the predicted parameters. The results of these experiments are then analyzed to determine their accuracy and consistency with the predicted outcomes. \cite{ref57, ref58} 

Any discrepancies or areas for improvement identified during the result analysis phase led to the refinement of the predictive models and experimental protocols. The feedback loop is a critical component for this approach, ensuring that each iteration of the process incorporates learnings from previous experiments. This continuous improvement cycle enhances the accuracy and efficiency of the self-driving laboratory, ultimately leading to the precise synthesis of materials with desired properties. Thus, the implementation of SciQu underscores its potential to revolutionize materials research, providing a scalable, efficient, and accurate approach to material synthesis and property prediction.

\section{Conclusion}
SciQu demonstrates a significant advancement in materials science by integrating machine learning with automated data extraction to predict and optimize material properties by addressing the inefficiencies and biases of traditional methods. By leveraging AI and NLP technologies, SciQu streamlines the literature review process and enhances the precision of material property predictions, such as the refractive index, with a notable RMSE of $\sim$ 0.068 and $R^{2}$ $\sim$ 0.94. This tool offers a robust solution for self-driving laboratories, which not only accelerates research but also improves the accuracy and reproducibility of experimental outcomes, marking a pivotal step towards fully autonomous materials discovery and optimization.

\section{Computational details}
\subsection{Building the SciQu for Automated Data Mining}
The SciQu tool allows users to upload PDF files and query their contents for relevant information. The application is built using Streamlit for the user interface and several LangChain components for backend processing. Upon uploading a PDF, the file is processed using the UnstructuredPDFLoader to extract its text. The extracted text is then split into chunks using the RecursiveCharacterTextSplitter with a chunk size of 700 and an overlap of 100. These chunks are embedded using OllamaEmbeddings and stored in a Chroma vector database. Users can ask questions related to the uploaded document through a text input field. The questions are processed using a MultiQueryRetriever, which generates multiple perspectives of the query to enhance retrieval accuracy. The retrieved context is then passed to a ChatOllama model to generate responses based on the given context. The responses are displayed to the user, and the query-answer pairs are saved in the session state for history tracking. The application supports dynamic user interaction and aims to improve the efficiency and accuracy of literature review by automating data extraction and query handling.\\
As a pilot project, we have utilized the 20 materials and their properties as input descriptors for predicting the refractive index (Table \ref{table:Table1}); the order of the materials are $K_{2}Te$, $K_{2}O$, BaS, $Na_{2}Te$, SnSe, CaS, MgS, $CdI_{2}$, $CdBr_{2}$, YN, $HgF_{2}$, SnO, BN, $PtO_{2}$, $K_{2}S$, BeS, $MgI_{2}$, RbBr, $VCl_{2}$, $Na_{2}S$, respectively.

\subsection{Integration of the ML training with the SciQu}
The demonstration of the SciQu training has been done by using 20 materials. After installing the necessary libraries, the dataset is loaded, and selected attributes are extracted, including refractive index, band gap, ferroelectricity, and others. The dataset is then checked for any missing values. Input features (X) and the target variable (y) are defined, selecting relevant columns. The data is split into training and testing sets using a 70-30 split. A Random Forest Regressor model with 100 estimators is created and trained on the training data. Predictions are made on the test set, and the model's performance is evaluated using Root Mean Squared Error (RMSE) and R-squared score. Finally, regression and residual plots are generated using Seaborn to visualize the model's performance.

\section*{Code Availability}
The code utilized for building the SciQu has been available on the GitHub repository (https://github.com/ABnano/SciQu).

\section*{Acknowledgment}
The author expresses deep gratitude to Prof. Dipankar Mandal at INST Mohali for insightful discussions and guidance. Special thanks are extended to Dr. Sahil Dani at Chitkara University for his valuable feedback.The author also acknowledges the University Grants Commission (UGC) for providing the fellowship (1354/CSIR-UGC NET DEC. 2018). Additionally, sincere appreciation is extended to Param Smriti for offering the high-performance computing facilities essential for this work.

\bibliographystyle{unsrt}
\bibliography{references}  

\begin{thebibliography}{10}

\bibitem{ref1}
Liliana Marinescu, Denisa Ficai, Ovidiu Oprea, Alexandru Marin, Anton Ficai, Ecaterina Andronescu, and Alina-Maria Holban.
\newblock Optimized synthesis approaches of metal nanoparticles with antimicrobial applications.
\newblock {\em Journal of Nanomaterials}, 2020(1):6651207, 2020.

\bibitem{ref2}
Emory~M. Chan, Chenjie Xu, Alexander Mao, Gang Han, Jonathan~S. Owen, Bruce~E. Cohen, and Delia~J. Milliron.
\newblock Reproducible, high-throughput synthesis of colloidal nanocrystals for optimization in multidimensional parameter space.
\newblock {\em Nano Letters}, 10(5):1874--1885, 2010.

\bibitem{ref3}
Solomon~A. Akintelu, Bin Yao, and Abayomi~S. Folorunso.
\newblock A review on synthesis, optimization, mechanism, characterization, and antibacterial application of silver nanoparticles synthesized from plants.
\newblock {\em Journal of Chemistry}, 2020:1--12, 2020.

\bibitem{ref4}
Anand Babu, Sanghmitra~G. Ghatnekar, Ashish Saxena, and Debabrata Mandal.
\newblock Can entanglement-enhanced quantum kernels improve data classification?
\newblock {\em ArXiv}, 2024.
\newblock arXiv:2406.01948.

\bibitem{ref5}
Sweta Kumari, Seema Raturi, Shikha Kulshrestha, Kavita Chauhan, Sudhir Dhingra, Károly András, and Tej Singh.
\newblock A comprehensive review on various techniques used for synthesizing nanoparticles.
\newblock {\em Journal of Materials Research and Technology}, 27:1739--1763, 2023.

\bibitem{ref6}
Farid Mekki-Berrada, Zheyu Ren, Hongyu Tan, Wai~Kwan Wong, Zhi Fang, Jun Xie, and Xuewen Wang.
\newblock Two-step machine learning enables optimized nanoparticle synthesis.
\newblock {\em NPJ Computational Materials}, 7(1), 2021.

\bibitem{ref7}
Oleksandr Voznyy, Lev Levina, Jun Fan, Marina Askerka, Alok Jain, Mark Choi, and Edward~H. Sargent.
\newblock Machine learning accelerates discovery of optimal colloidal quantum dot synthesis.
\newblock {\em ACS Nano}, 13(10):11122--11128, 2019.

\bibitem{ref8}
Yuyu Jia, Xin Hou, Zhongli Wang, and Xun Hu.
\newblock Machine learning boosts the design and discovery of nanomaterials.
\newblock {\em ACS Sustainable Chemistry \& Engineering}, 9(18):6130--6147, 2021.

\bibitem{ref9}
Anand Babu and Debabrata Mandal.
\newblock Roadmap to human–machine interaction through triboelectric nanogenerator and machine learning convergence.
\newblock {\em ACS Applied Energy Materials}, 7(3):822--833, 2024.

\bibitem{ref10}
Milad Abolhasani and Eugenia Kumacheva.
\newblock The rise of self-driving labs in chemical and materials sciences.
\newblock {\em Nature Synthesis}, 2(6):483--492, 2023.

\bibitem{ref11}
Dmitry~A. Boiko, Ross MacKnight, Brendan Kline, and Gustavo Gomes.
\newblock Autonomous chemical research with large language models.
\newblock {\em Nature}, 624(7992):570--578, 2023.

\bibitem{ref12}
Martin Seifrid, Robert Pollice, Agustin Aguilar-Granda, Ziyun~M. Chan, Kei Hotta, Choon-Tak Ser, and Alán Aspuru-Guzik.
\newblock Autonomous chemical experiments: Challenges and perspectives on establishing a self-driving lab.
\newblock {\em Accounts of Chemical Research}, 55(17):2454--2466, 2022.

\bibitem{ref13}
Bruce~P. MacLeod, Fraser G.~L. Parlane, T.~Dylan Morrissey, Florian Häse, Loïc~M. Roch, Kevan~E. Dettelbach, and Curtis~P. Berlinguette.
\newblock Self-driving laboratory for accelerated discovery of thin-film materials.
\newblock {\em Science Advances}, 6(20), 2020.

\bibitem{ref14}
Florian Häse, Loïc~M. Roch, and Alán Aspuru-Guzik.
\newblock Next-generation experimentation with self-driving laboratories.
\newblock {\em Trends in Chemistry}, 1(3):282--291, 2019.

\bibitem{ref15}
Steven~G. Baird and Taylor~D. Sparks.
\newblock What is a minimal working example for a self-driving laboratory?
\newblock {\em Matter}, 5(12):4170--4178, 2022.

\bibitem{ref16}
Naoko Yoshikawa, Miltiadis Skreta, Kiavash Darvish, Sergio Arellano-Rubach, Zhijun Ji, Lars~B. Kristensen, and Abhishek Garg.
\newblock Large language models for chemistry robotics.
\newblock {\em Autonomous Robots}, 47(8):1057--1086, 2023.

\bibitem{ref17}
Sebastian~D. Rihm, Jing Bai, Aleksandar Kondinski, Sebastian Mosbach, John Akroyd, and Markus Kraft.
\newblock Transforming research laboratories with connected digital twins.
\newblock {\em Nexus}, 1(1):100004, 2024.

\bibitem{ref18}
Muhammad Murtaza, Cheng Cheng, Mohammad Fard, and John Zeleznikow.
\newblock Transforming driver education: A comparative analysis of llm-augmented training and conventional instruction for autonomous vehicle technologies.
\newblock {\em International Journal of Artificial Intelligence in Education}, 2024.

\bibitem{ref19}
Fernando Delgado-Licona and Milad Abolhasani.
\newblock Research acceleration in self-driving labs: Technological roadmap toward accelerated materials and molecular discovery.
\newblock {\em Advanced Intelligent Systems}, 5(4), 2022.

\bibitem{ref20}
Mikhail~A. Soldatov, Viktoria~V. Butova, Dmitry~M. Pashkov, Marina~A. Butakova, Pavel V. Chernov Andrey~V. Medvedev, and Soldatov~Alexander V.
\newblock Self-driving laboratories for development of new functional materials and optimizing known reactions.
\newblock {\em Nanomaterials}, 11(3):619, 2021.

\bibitem{ref21}
{Semantic Scholar Open Research Corpus}.
\newblock S2orc: The semantic scholar open research corpus, 2020.

\bibitem{ref22}
Iz~Beltagy, Kyle Lo, and Arman Cohan.
\newblock Scibert: A pretrained language model for scientific text.
\newblock {\em ArXiv}, 2019.

\bibitem{ref23}
Anand Chaudhari, Chaitanya Guntuboina, Hsin-Yu Huang, and Amir~Barati Farimani.
\newblock Alloybert: Alloy property prediction with large language models.
\newblock {\em ArXiv}, 2024.

\bibitem{ref24}
Swarnadeep Agarwal, Issa~H. Laradji, Laurent Charlin, and Chris Pal.
\newblock Litllm: A toolkit for scientific literature review.
\newblock {\em ArXiv}, 2024.

\bibitem{ref25}
Jan Lála, Owen O'Donoghue, Andrei Shtedritski, Sam Cox, Samuel~G. Rodriques, and Andrew~D. White.
\newblock Paperqa: Retrieval-augmented generative agent for scientific research.
\newblock {\em ArXiv}, 2023.

\bibitem{ref26}
Harold Hysmith, Eileen Foadian, Sushree~P. Padhy, Sergei~V. Kalinin, Robert~G. Moore, Olga~S. Ovchinnikova, and Masoud Ahmadi.
\newblock The future of self-driving laboratories: From human in the loop interactive ai to gamification.
\newblock 2024.

\bibitem{ref27}
George Tom, Stefanie~P. Schmid, Steven~G. Baird, Yuxin Cao, Kiavash Darvish, Hang Hao, and et~al.
\newblock Self-driving laboratories for chemistry and materials science, 2024.
\newblock https://doi.org/10.26434/chemrxiv-2024-rj946-v2.

\bibitem{ref28}
Ke~Chen, Yang Li, Weifeng Zhang, Yu~Liu, Peng Li, Rui Gao, Lian Hong, Mingzhu Tian, Xia Zhao, Zhenfeng Li, Dit-Yan Yeung, Hongyang Lu, and Xiantong Jia.
\newblock Automated evaluation of large vision-language models on self-driving corner cases.
\newblock {\em ArXiv}, 2024.

\bibitem{ref29}
Dong Fu, Wei Lei, Liguo Wen, Peibin Cai, Siyuan Mao, Ming Dou, Bicheng Shi, and Yuning Qiao.
\newblock Limsim++: A closed-loop platform for deploying multimodal llms in autonomous driving.
\newblock {\em ArXiv}, 2024.

\bibitem{ref30}
Yuchen Wei, Zijun Wang, Yixuan Lu, Chang Xu, Cheng Liu, Haoyang Zhao, Sibo Chen, and Yafei Wang.
\newblock Editable scene simulation for autonomous driving via collaborative llm-agents.
\newblock {\em ArXiv}, 2024.

\bibitem{ref31}
Yuzhuo Cui.
\newblock Charting the path toward full autonomous driving with large language models.
\newblock {\em IEEE Transactions on Intelligent Vehicles}, 9(1):1450--1464, 2024.

\bibitem{ref32}
Srinivas~P. Sharan, Francesco Pittaluga, V.~Kamakshidhar K, and Manmohan Chandraker.
\newblock Llm-assist: Enhancing closed-loop planning with language-based reasoning.
\newblock {\em ArXiv}, 2023.

\bibitem{ref33}
E.~Zintl, A.~Harder, and B.~Dauth.
\newblock Gitterstruktur der oxide, sulfide, selenide und telluride des lithiums, natrums und kaliums.
\newblock {\em Zeitschrift für Elektrochemie}, 40:588--593, 1934.

\bibitem{ref34}
Yurii~N. Zhuravlev and Olga~S. Obolonskaya.
\newblock Structure, mechanical stability and chemical bond in alkali metal oxides.
\newblock {\em Journal of Structural Chemistry}, 51:1005--1013, 2010.

\bibitem{ref35}
Jean Flahaut, Lucien Domange, and Marcel Patrie.
\newblock Combinaisons formes par les sulfures des elements du groupe des terres rares. etude cristallographique des phases ayant le type structural du phosphure de thorium th3 p4.
\newblock {\em Bulletin de la Société Chimique de France}, pages 2048--2054, 1962.

\bibitem{ref36}
W.~Klemm, H.~Sodomann, and P.~Langmesser.
\newblock Beitraege zur kenntnis der alkalimetallchalkogenide.
\newblock {\em Zeitschrift für Anorganische und Allgemeine Chemie}, 241:281--304, 1939.

\bibitem{ref37}
Lev~S. Palatnik and Valery~V. Levitin.
\newblock X-ray investigation of alloys sn-se, zn-se, cd-se, and ag-se.
\newblock {\em Doklady Akademii Nauk SSSR}, 96:975--978, 1954.

\bibitem{ref38}
Takaaki Kobayashi, Kazuo Susa, and Satoshi Taniguchi.
\newblock Preparation and semiconductive properties of rock salt type solid solution systems, cd1-x mx s (m = sr, ca, mg, pb, sn).
\newblock {\em Journal of Physics and Chemistry of Solids}, 40:781--785, 1979.

\bibitem{ref39}
Syed~M. Alay-e Abbas and Ameer Shaukat.
\newblock Fp-lapw calculations of structural, electronic, and optical properties of alkali metal tellurides: M2 te (m : Li, na, k, and rb).
\newblock {\em Journal of Materials Science}, 46:1027--1037, 2011.

\bibitem{ref40}
David~T. Hodul and Angelica~M. Stacy.
\newblock The structure and electronic properties of the solid solutions (zrxti1-x)1+ys2.
\newblock {\em Journal of Solid State Chemistry}, 62:328--334, 1986.

\bibitem{ref41}
Syed~M. Alay-e Abbas, Naveed Sabir, Yasir Saeed, and Ameer Shaukat.
\newblock Electronic and optical properties of alkali metal selenides in anti-caf2 crystal structure from first-principles.
\newblock {\em Journal of Alloys and Compounds}, 503:10--18, 2010.

\bibitem{ref42}
Jürgen Peters and Bernt Krebs.
\newblock Silicon disulphide and silicon diselenide: A reinvestigation.
\newblock {\em Acta Crystallographica Section B: Structural Crystallography and Crystal Chemistry}, 38:1270--1272, 1982.

\bibitem{ref43}
Kazuo Susa, Takaaki Kobayashi, and Satoshi Taniguchi.
\newblock High-pressure synthesis of rock-salt type cds using metal sulfide additives.
\newblock {\em Journal of Solid State Chemistry}, 33:197--202, 1980.

\bibitem{ref44}
Walter~J. Moore and Linus Pauling.
\newblock The crystal structures of the tetragonal monoxides of lead, tin, palladium, and platinum.
\newblock {\em Journal of the American Chemical Society}, 63:1392--1394, 1941.

\bibitem{ref45}
Gao Shangpeng.
\newblock Crystal structures and band gap characters of h-bn polytypes predicted by the dispersion corrected dft and gw method.
\newblock {\em Solid State Communications}, 152:1817--1820, 2012.

\bibitem{ref46}
James~R. McBride, George~W. Graham, Chester~R. Peters, and William~H. Weber.
\newblock Growth and characterization of reactively sputtered thin-film platinum oxides.
\newblock {\em Journal of Applied Physics}, 69:1596--1604, 1991.

\bibitem{ref47}
Charles~D. West.
\newblock The crystal structures of some alkali hydrosulfides and monosulfides.
\newblock {\em Zeitschrift für Kristallographie}, 88:97--115, 1934.

\bibitem{ref48}
Khadidja Hacini, Salim Ghemid, Hacene Meradji, and Fouad El~Haj~Hassan.
\newblock Theoretical study of structural, electronic, and thermal properties of zn1-x bex s ternary alloy.
\newblock {\em Computational Materials Science}, 50:3080--3084, 2011.

\bibitem{ref49}
Megan~A. Brogan, Alexander~J. Blake, Claire Wilson, and Duncan~H. Gregory.
\newblock Magnesium diiodide, mgi2.
\newblock {\em Acta Crystallographica Section C: Structural Chemistry}, 59:i136--i138, 2003.

\bibitem{ref50}
Walter~P. Davey.
\newblock Cubic form of certain ions determined by x-ray structure analysis.
\newblock {\em Physical Review}, 17:402--403, 1921.

\bibitem{ref51}
Kazumichi Hirakawa, Hiroshi Kadowaki, and Kozo Ubukoshi.
\newblock Study of frustration effects in two-dimensional triangular lattice antiferromagnets - neutron powder diffraction study of vx2, x = cl, br, and i.
\newblock {\em Journal of the Physical Society of Japan}, 52:1814--1824, 1983.

\bibitem{ref52}
Philippe~R. Bonneau, Roy~F. Jarvis~Jr., and Richard~B. Kaner.
\newblock Solid-state metathesis as a quick route to transition-metal mixed dichalcogenides.
\newblock {\em Inorganic Chemistry}, 31:2127--2132, 1992.

\bibitem{ref53}
Ivo~W. Bassi, Fabio Polato, Monica Calcaterra, and Jan C.~J. Bart.
\newblock A new layer structure of mgcl2 with hexagonal close packing of the chlorine atoms.
\newblock {\em Zeitschrift für Kristallographie}, 159:297--302, 1982.

\bibitem{ref54}
Jin Wu, Chao Lu, Alejandra Arrieta, Tianci Yue, and Syed Ali.
\newblock Reality bites: Assessing the realism of driving scenarios with large language models.
\newblock In {\em Proceedings of the 2024 IEEE/ACM First International Conference on AI Foundation Models and Software Engineering (FORGE '24)}, pages 40--51, New York, NY, USA, 2024. Association for Computing Machinery.

\bibitem{ref55}
Xiang Tian, Jiahao Gu, Bingzhe Li, Yijie Liu, Yifan Wang, Zichen Zhao, Kai Zhan, Pengfei Jia, Xiang Lang, and Han Zhao.
\newblock Drivevlm: The convergence of autonomous driving and large vision-language models.
\newblock {\em ArXiv}, 2024.

\bibitem{ref56}
Santiago Miret and Nandan~M. Krishnan.
\newblock Are llms ready for real-world materials discovery?
\newblock {\em ArXiv}, 2024.

\bibitem{ref57}
Rui Zhang, Xingyi Guo, Wei Zheng, Chiyu Zhang, Kurt Keutzer, and Liang Chen.
\newblock Instruct large language models to drive like humans.
\newblock {\em ArXiv}, 2024.

\bibitem{ref58}
Yiping Zhou, Linghao Huang, Qunxiang Bu, Jiewei Zeng, Ting Li, Han Qiu, Hongping Zhu, Min Guo, Yuning Qiao, and Huizi Li.
\newblock Embodied understanding of driving scenarios.
\newblock {\em ArXiv}, 2024.

\end{thebibliography}






\end{document}